\documentclass[twocolumn,showpacs,aps,pra]{revtex4-1}

\usepackage{amsmath,amsfonts,amssymb}
\usepackage{wrapfig}
\usepackage{graphicx}
\usepackage{bbm}
\usepackage[normalem]{ulem}
\usepackage{xcolor}

\begin{document}

\title{Securing quantum networking tasks\\ with multipartite Einstein-Podolsky-Rosen steering}\date{\today}

\author{Chien-Ying Huang$^{1,2}$}
\author{Neill Lambert$^{3}$}
\author{Che-Ming Li$^{1}$}
\email{cmli@mail.ncku.edu.tw}
\author{Yen-Te Lu$^{1}$}
\author{Franco Nori$^{3,4}$}

\affiliation{$^1$Department of Engineering Science, National Cheng Kung University, Tainan 701, Taiwan}
\affiliation{$^2$Graduate Institute of Photonics and Optoelectronics, National Taiwan University, Taipei 10617, Taiwan}
\affiliation{$^3$Theoretical Quantum Physics Laboratory, RIKEN Cluster for Pioneering Research, Wako-shi, Saitama 351-0198, Japan}
\affiliation{$^4$Department of Physics, University of Michigan, Ann Arbor, Michigan 48109-1040, USA}

\begin{abstract}
Einstein-Podolsky-Rosen (EPR) steering is the explicit demonstration of the fact that the measurements of one party can influence the quantum state held by another, distant, party, and do so even if the measurements themselves are untrusted. This has been shown to allow one-sided device-independent quantum-information tasks between two remote parties. However, in general, advanced multiparty protocols for generic quantum technologies, such as quantum secret sharing and blind quantum computing for quantum networks, demand multipartite quantum correlations of graph states shared between more than two parties. Here, we show that, when one part of a quantum multidimensional system composed of a two-colorable graph state (e.g., cluster and Greenberger-Horne-Zeilinger states) is attacked by an eavesdropper using a universal cloning machine, only one of the copy subsystems can exhibit multipartite EPR steering \emph{but not both}. Such a no-sharing restriction secures both state sources and channels against cloning-based attacks for generic quantum networking tasks, such as distributed quantum-information processing, in the presence of uncharacterized measurement apparatuses.
\end{abstract}

\maketitle

\section{Introduction}

Einstein-Podolsky-Rosen (EPR) steering \cite{Schrodinger35} is a unique part of EPR non-locality \cite{EPR35}. It determines which
states can be remotely prepared at one location, by performing a measurement at another. Since its operational definition introduced by \cite{Wiseman07}, this ``spooky action at a distance" appears to be a subtle form of quantum correlation intermediate between entanglement and Bell non-locality. Recently, this operational formulation has been utilized to exploit EPR steering to perform quantum key distribution \cite{Branciard12} even if one party's measurement devices are untrusted. Quantum information, however, involves many other types of applications. Indeed, in general, many quantum-information tasks inevitably require transmitting, sharing or processing quantum information between more than two spatially separated quantum nodes, representing separated quantum systems, via quantum channels \cite{Kimble08,Duan10,Northup14,Reiserer15}, which together form distributed quantum networks.

\begin{figure}[t]
\includegraphics[width=8.8cm,angle=0]{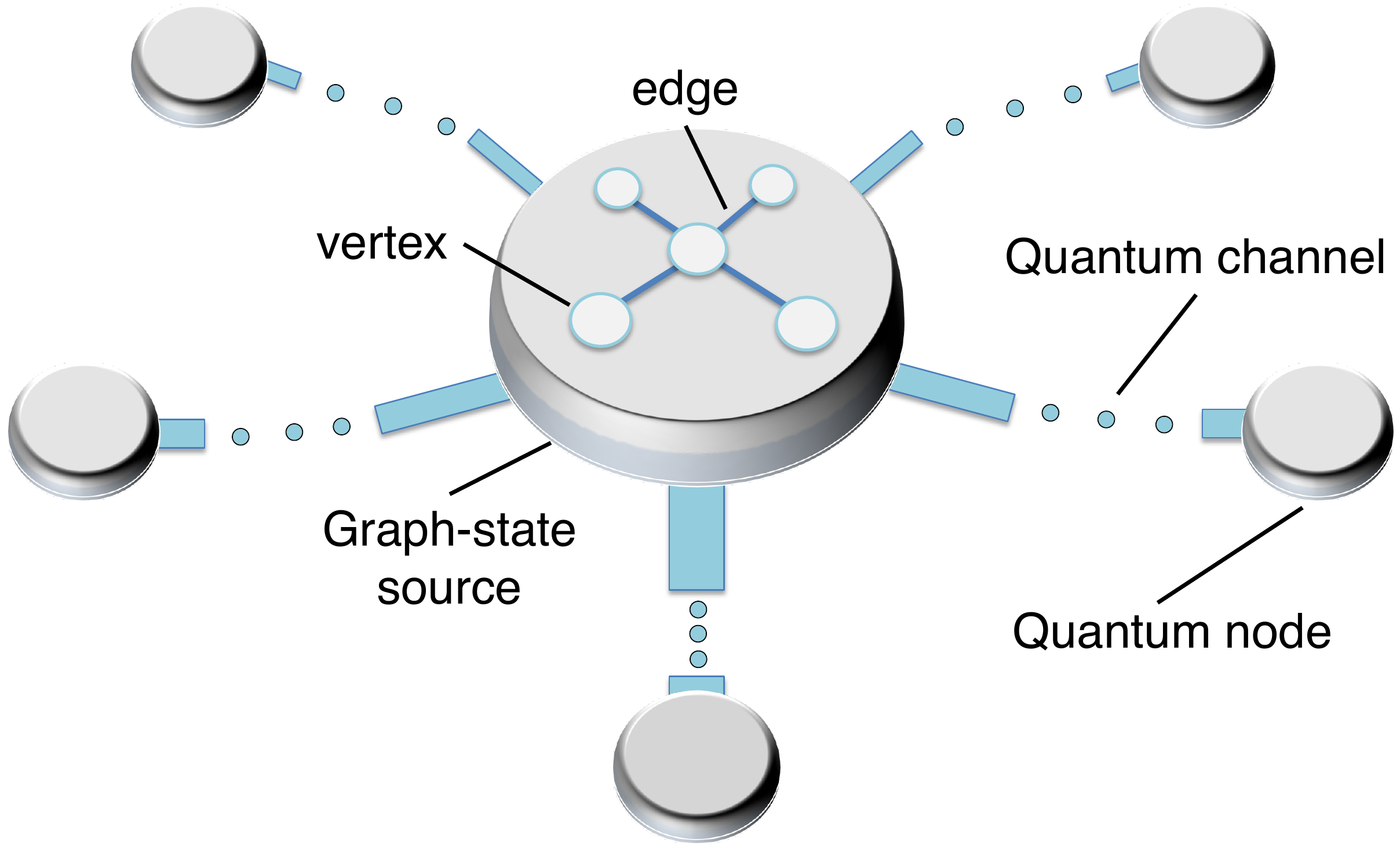}
\caption{(Color online). Graph states for quantum networks. Graph states are created from a graph-state source, such as the star-graph states shown here, with the goal of using them for quantum-information tasks. Each qudit is sent from the source to the corresponding quantum node through a quantum channel to implement said task, such as QSS \cite{Hillery99,Chen05,Markham08,Bell14a} or MBQC \cite{Raussendorf01,Raussendorf03,Walther05,Chen07,Briegel09}. Whereas, for BQC \cite {Broadbent09,Barz12}, a specific initial state is sent from quantum nodes to the source for creating blind graph states. This construction is the essence of the modular and plug-and-play quantum network architecture \cite{Pirker17}.} \label{GraphNet}
\end{figure}

The multipartite quantum correlations present in graph states \cite{Hein04,Hein05} are thought to act as an important resource; a type of fuel that powers a wide range of quantum strategies and protocols for networking tasks, including distributed quantum-information processing, such as quantum secret sharing (QSS) \cite{Hillery99,Chen05,Markham08,Bell14a}, universal measurement-based quantum computation (MBQC) \cite{Raussendorf01,Raussendorf03,Walther05,Chen07,Briegel09}, quantum error correction codes (QECC) \cite{Schlingemann01,Looi2008,Bell14b}, and blind quantum computing (BQC) \cite {Broadbent09,Barz12}. Quantum metrology takes advantage of this fuel as well, to offer higher precision than classical methods \cite{Giovannetti11}, such as a quantum network of clocks \cite{Komar14}. Graph states are even used to establish the basic building blocks for general modular architectures of quantum networks \cite{Pirker17}. An $N$-qudit (quantum $d$-dimensional systems) graph state can be represented by a graph $G(V,E)$ \cite{Hein04,Hein05,Raussendorf03,Zhou03}. In general, the graph $G$ comprises the vertex set $V$ with a cardinality $|V|=N$, representing the qudits, and the edge set $E$ each of which joins two vertices, representing interacting pairs of qudits; see Fig.~\ref{GraphNet}. If the vertices of the graph $G$ can be divided into $q$ sets, and the vertices of each set are given a color, such that adjacent vertices have different colors, then the graph is called a $q$-colorable graph \cite{Hein04,Hein05}. See Appendix A.1 for a detailed illustration of graph states for $q=2$.

Compared with the many broad formulations, and potential applications, of quantum technologies, there is, so far, only a very preliminary conceptual understanding of multipartite EPR steering \cite{He13}. While detecting the steerability of multipartite systems \cite{Armstrong15,Cavalcanti15,Deng17} and genuine multipartite EPR steering \cite{He13,Li15} is possible, the fundamental issue of the role of such high-order EPR steering in securing quantum-information processing involving multiple participants remains unclear. Very recently, for relatively small numbers of participants, genuine tripartite steering for pure three-mode Gaussian states \cite{Xiang17} was shown to empower a partially device-independent QSS protocol \cite{Kogias17}. Moreover, many entangled systems in graph states have been created and manipulated coherently in various experimental implementations \cite{Monz11,Barends14,Kelly15,Wang16,Wang18}. The technological challenges facing the eventual realization of quantum technology suggests that they will inevitably rely on uncharacterized facilities and involve partially untrusted participants. While verification protocols of multipartite entanglement in the presence of untrusted parties, based on entanglement witness, have been proposed, these protocols are task-oriented and currently limited to Greenberger-Horne-Zeilinger (GHZ) states \cite{Pappa12,McCutcheon16}. On the other hand, according to the operational definition of EPR steering \cite{Wiseman07}, verifying steerability not only assures that the particles shared with the trusted parties are truly entangled but also excludes the presence of untrusted participant in the tasks. Therefore, such a physical model of steering can be utilized in the context of the trusted-untrusted-participant scenario of various distributed quantum-information tasks. However, does the higher-order EPR steerability of graph states preserve the security of quantum-information processing in such imperfect circumstances?

Here, in order to tackle the issue of security of quantum networks, we reveal the no-sharing of multipartite EPR steering for any two-colorable graph states \cite{Hein04,Hein05,Li10}, which include cluster states and GHZ states (i.e., star-graph states) as prominent illustrations of this class of graph states for quantum technologies \cite{Hillery99,Chen05,Markham08,Bell14a,Raussendorf01,Raussendorf03,Walther05,Chen07,Briegel09,Schlingemann01,Hein05,Looi2008,Bell14b,Broadbent09,Barz12,Giovannetti11,Komar14,Pirker17, Tanamoto06,You07,Tanamoto09}. Our scenario for identifying such no-sharing characteristics requires only the minimum of two local measurement settings for each quantum node, and can be applied to general quantum-information protocols based on normal local operations and classical communication, for instance, QSS, MBQC, and BQC, mentioned earlier.

We start with a definition of graph states in Schmidt form. Details of ideal graph states and the Schmidt decomposition are provided in Appendix A. In Sec. II we then define multipartite steering and introduce a measurable criterion for the presence of steering in such states based on mutual information. In appendix B we provide a concrete example, a complete derivation of the criterion in the Schmidt bases, and its tolerance to noise. In Sec. III we use this criterion to show that multipartite steering cannot be shared by a cloning machine, and thus its observation can be used to verify the security of a network, which is explicitly derived in Appendix C. Concrete examples based on QSS, MBQC, and BQC are illustrated in Sec. IV. Moreover, the steering is shown to set a lower bound on the key rate in the problem of QSS with a complete derivation provided in Appendix D. In Sec. V we finish with the implications of our results for general quantum networking tasks that demand two-colorable graph states \cite{Hillery99,Chen05,Markham08,Bell14a,Raussendorf01,Raussendorf03,Walther05,Chen07,Briegel09,Schlingemann01,Hein05,Looi2008,Bell14b,Broadbent09,Barz12,Giovannetti11,Komar14,Pirker17}. Insights and outlook of our work have been summarized in Sec. VI.

We assume that, after being created from a graph-state source (Fig.~\ref{GraphNet}), $N$ qudits (with dimension $d$) of the two-colorable graph state $\left|G_{2}\right\rangle$ are individually sent to $N$ parties of quantum nodes. In the trusted-untrusted-participant scenario, the $N$ parties are divided into two groups, say $A_{s}$ and $B_{s}$. With respect to this given bipartition, $A_{s}$ and $B_{s}$ can perceive that the state $\left|G_{2}\right\rangle$ connects them together through correlations between qudits, as described by the Schmidt form \cite{Li10}
\begin{equation}
\left|G_{2}\right\rangle=\frac{1}{\sqrt{d}}\sum_{v=0}^{d-1}\left|v\right\rangle_{A_{s}m}\otimes\left|v\right\rangle_{B_{s}m},\label{sch}
\end{equation}
where $\text{A}^{(S)}_{sm}=\{\left|v_{Am}\right\rangle_{A_{s}m}|v_{Am}\in\textbf{v}\}$ and $\text{B}^{(S)}_{sm}=\{\left|v_{Bm}\right\rangle_{B_{s}m}|v_{Bm}\in\textbf{v}\}$ are the orthonormal bases for $A_{s}$'s and $B_{s}$'s qudits with $m=1,2$ for two different Schmidt bases, respectively, and $\textbf{v}=\{0,1,...,d-1\}$. Since there are $d$ nonvanishing terms in the Schmidt form, the Schmidt rank \cite{Terhal2000} of the graph state is $d$. See Appendix A.2 for the derivation of the Schmidt decomposition (\ref{sch}).

\section{EPR steering between multi-quantum nodes}

In order to concretely represent the multipartite EPR steering of the state $\left|G_{2}\right\rangle$ (\ref{sch}), we consider a general model to describe states in the presence of uncharacterized measurement apparatuses. In our scenario, two possible measurements can be performed on each particle ($m_{k}=1,2$ for the $k$th particle), and each local measurement has $d$ possible outcomes, $v_{k}^{(m_{k})}\in\textbf{v}$. That is, each party can implement quantum measurements of observables with the nondegenerate eigenvectors $\{\left|0\right\rangle_{k,1}=\left|0\right\rangle_{k},\left|1\right\rangle_{k,1}=\left|1\right\rangle_{k}, ..., \left|d-1\right\rangle_{k,1}=\left|d-1\right\rangle_{k}\}$ for $m_{k}=1$ and $\{\left|0\right\rangle_{k,2}=\hat{\text{F}}_{k}\left|0\right\rangle_{k},\left|1\right\rangle_{k,2}=\hat{\text{F}}_{k}\left|1\right\rangle_{k}, ..., \left|d-1\right\rangle_{k,2}=\hat{\text{F}}_{k}\left|d-1\right\rangle_{k}\}$ for $m_{k}=2$, where $\hat{\text{F}}_{k}$ is the quantum Fourier transformation (see detailed definition in Appendix A.1). We assume that the measurement devices used by the parties in $A_{s}$ are uncharacterized, i.e., untrustworthy. In the worst case, $A_{s}$'s measurement outcomes may be randomly generated from the measurement apparatuses themselves. In general, an unqualified source of the graph state, or noisy channels, may also lead to the same effect. Classical simulations, under the assumption of realism, can then describe the measurement results of $A_{s}$, which empowers $A_{s}$'s ability to mimic the target state according to classical realism. Such an ability makes EPR steerability a strictly stronger quantum correlation than entanglement, which corresponds to the trusted-trusted-participant scenario in terms of operational definitions. In this case, with respect to a given bipartite splitting of the $N$ parties, say $\alpha$, the final state of the $N$ particles can be specified by classical realistic theories, which predict that the particles are in a state belonging to a fixed set
\begin{equation}
\{v_{k}^{(1)},v_{k}^{(2)},\lambda_{\alpha}|\forall k\in \textbf{a}_{s}\},\label{vrealism}
\end{equation}
where the random variable $\lambda_{\alpha}$ corresponds to an unknown quantum state $\rho _{\lambda_{\alpha}}$ shared by the parties in $B_{s}$, and $\textbf{a}_{s}$ denotes the indexing set for the parties in $A_{s}$.

The final states of the $N$-particle system may depend on unknown sources of randomness from the measurement apparatuses, graph-state source or channels, such that the above deterministic scenario becomes a probabilistic one. For a given bipartition $\alpha$, they can then be characterized by the state probabilities, $P_{\alpha}(v_{k}^{(1)},v_{k}^{(2)},\lambda_{\alpha}|\forall k\in \textbf{a}_{s})$, to be
\begin{eqnarray}
\rho_{B_{s}}=&&\sum_{v_{k}^{(1)},v_{k}^{(2)}}\sum_{\lambda_{\alpha}} P_{\alpha}(v_{k}^{(1)},v_{k}^{(2)},\lambda_{\alpha}|\forall k\in \textbf{a}_{s})}\;\rho_{\lambda_{\alpha}.\label{rhobs}
\end{eqnarray}
If a state can offer stronger correlations between $A_{s}$ and $B_{s}$ than any strategies involving $\rho_{B_{s}}$ for all possible $\lambda_{\alpha}$, which can be explained by classical realistic theories in the presence of uncharacterized measurement apparatuses, we say that it possesses multipartite EPR steerability. Typically, intricate optimization processes are needed to characterize multipartite EPR steerability in (\ref{rhobs}), and thus a practical way to efficiently detect EPR steering for large scale quantum systems is still an open problem. Here in order to circumvent this difficulty, we utilize the Schmidt decomposition \cite{Li10} introduced above and the entropic uncertain relation \cite{Tomamichel11,Coles17} to propose an efficient criterion to verify multipartite EPR steerability.

For all two-colorable graph states, $\left|G_{2}\right\rangle$, there are steering correlations between $A_{s}$ and $B_{s}$ which cannot be mimicked by the states $\rho_{B_{s}}$~Eq.~(\ref{rhobs}). We first use the definition of information shared between $A_{s}$ and $B_{s}$ to certify that the non-classical mutual dependence between the results of $A_{s}$'s and $B_{s}$'s measurements on $\left|G_{2}\right\rangle$ is larger than the dependence of $B_{s}$'s measurement outcomes on the state $\rho_{B_{s}}$. Hence the ability for $A_{s}$ to steer $B_{s}$'s state is confirmed if the mutual dependence between the measurement results of $A_{s}$ and $B_{s}$ is stronger than the dependence of $B_{s}$'s measurement outcomes on the state $\rho_{B_{s}}$. This steering condition can be concretely represented in terms of the mutual information as follow,
\begin{equation}
I_{A_{s}B_{s}}\equiv\sum_{m=1}^{2}I_{A_{sm}B_{sm}}> \sum_{m=1}^{2}I_{\lambda_{\alpha}B_{sm}},\label{Is}
\end{equation}
where
\begin{equation}
I_{A_{sm}B_{sm}} =H_{m}(B_{s}) - H_{m}(B_{s}|A_{s}),\nonumber
\end{equation}
and
\begin{equation}
I_{\lambda_{\alpha}B_{sm}}=H_{m}(B_{s}) - H_{m}(B_{s}|\lambda_{\alpha}).\nonumber
\end{equation}
Such steering criterion generalizes the existing steering condition for two qudits \cite{Li15b,Chiu16} to multipartite systems. We assume that each particle is locally measured by its holder, and the parties, who implement quantum measurements, can perform positive operator valued measurements (POVMs) with sets of measurement operators composed of locally measurable operators, $\text{A}_{sm}$ and $\text{B}_{sm}$, that are extracted from and commutative with the elements of the Schmidt bases $\text{A}^{(S)}_{sm}$ and $\text{B}^{(S)}_{sm}$ \cite{Li10}. See Appendix B.1 for a detailed example. The measurement outcomes of $A_{s}$ and $B_{s}$: $\{a_{s,1},b_{s,1}\}$ and $\{a_{s,2},b_{s,2}\}$, are then obtained from the measurements which corresponds to $(\text{A}_{s1},\text{B}_{s1})$ and $(\text{A}_{s2},\text{B}_{s2})$, respectively. They are used to determine the entropy of $B_{s}$'s outcomes:
\begin{equation}
H_{m}(B_{s}) = - \sum_{b_{s,m}} P(b_{s,m}) \log_{2}P(b_{s,m}),\nonumber
\end{equation}
and the entropy conditioned on $A_{s}$'s results:
\begin{equation}
H_{m}(B_{s}|A_{s})=\sum_{a_{s,m}}P(a_{s,m}) H_{m}\left( B_{s} | a_{s,m} \right).\nonumber
\end{equation}

For the two-colorable graph states, $\left|G_{2}\right\rangle$, here we derive the overall correlation between $A_{s}$ and $B_{s}$ in terms of mutual information using corresponding quantum measurements, $\text{A}_{sm}$ and $\text{B}_{sm}$. The entropy of $B_{s}$'s outcomes are
\begin{equation}
H_{1}(B_{s}) = H_{2}(B_{s}) = \log_{2}d,\nonumber
\end{equation}
and the entropy conditioned on $A_{s}$'s results are
\begin{equation}
H_{1}(B_{s}|A_{s}) = H_{2}(B_{s}|A_{s}) = 0.\nonumber
\end{equation}
Therefore, the mutual information of $A_{s}$'s and $B_{s}$'s measurements become
\begin{equation}
I_{A_{s1}B_{s1}} = I_{A_{s2}B_{s2}} = \log_{2}d.\nonumber
\end{equation}
We can thus obtain,
\begin{equation}
I_{A_{s}B_{s}} = 2\log_{2}d. \label{MI}
\end{equation}
This result can be easily seen from the example of the 3-qubit star-graph state given in Appendix B.1. These two-colorable graph states held by trusted parties with perfect conditions are useful resources for a variety of quantum-information tasks, such as QSS \cite{Hillery99,Chen05,Markham08,Bell14a}, MBQC \cite{Raussendorf01,Raussendorf03,Walther05,Chen07,Briegel09}, and BQC \cite {Broadbent09,Barz12}.

In addition, since the unknown quantum state $\rho_{\lambda_{\alpha}}$ satisfies the entropic uncertainty relation under the two POVMs, $\text{B}_{s1}$ and $\text{B}_{s2}$, which are complementary to each other \cite{Tomamichel11,Coles17}:
\begin{equation}
H_{1}(B_{s}|\lambda_{\alpha}) + H_{2}(B_{s}|\lambda_{\alpha}) \geq \log_{2}d,\label{uncertain}
\end{equation}
the steering criterion~(\ref{Is}) then puts a bound on $I_{A_{s}B_{s}}$ as
\begin{equation}
I_{A_{s}B_{s}} > \log_{2}d.\label{Icondition}
\end{equation}
It is clear that, from the above derivation, the trusted and untrusted roles of $A_{s}$ and $B_{s}$ can be exchanged. Hence, given the knowledge of which group is trusted, the criterion (\ref{Icondition}) negates the possibility that either $A_{s}$'s or $B_{s}$'s measurement results can be classically simulated. As shown in Appendix B.2, the steering condition can also be described in the Schmidt bases $(\text{A}^{(S)}_{sm},\text{B}^{(S)}_{sm})$. Moreover, the criterion (\ref{Icondition}) is robust against white noise, independent of the number of participants, $N$. See the detailed discussions in Appendix B.3.

\begin{figure}[t]
\includegraphics[width=8.8cm,angle=0]{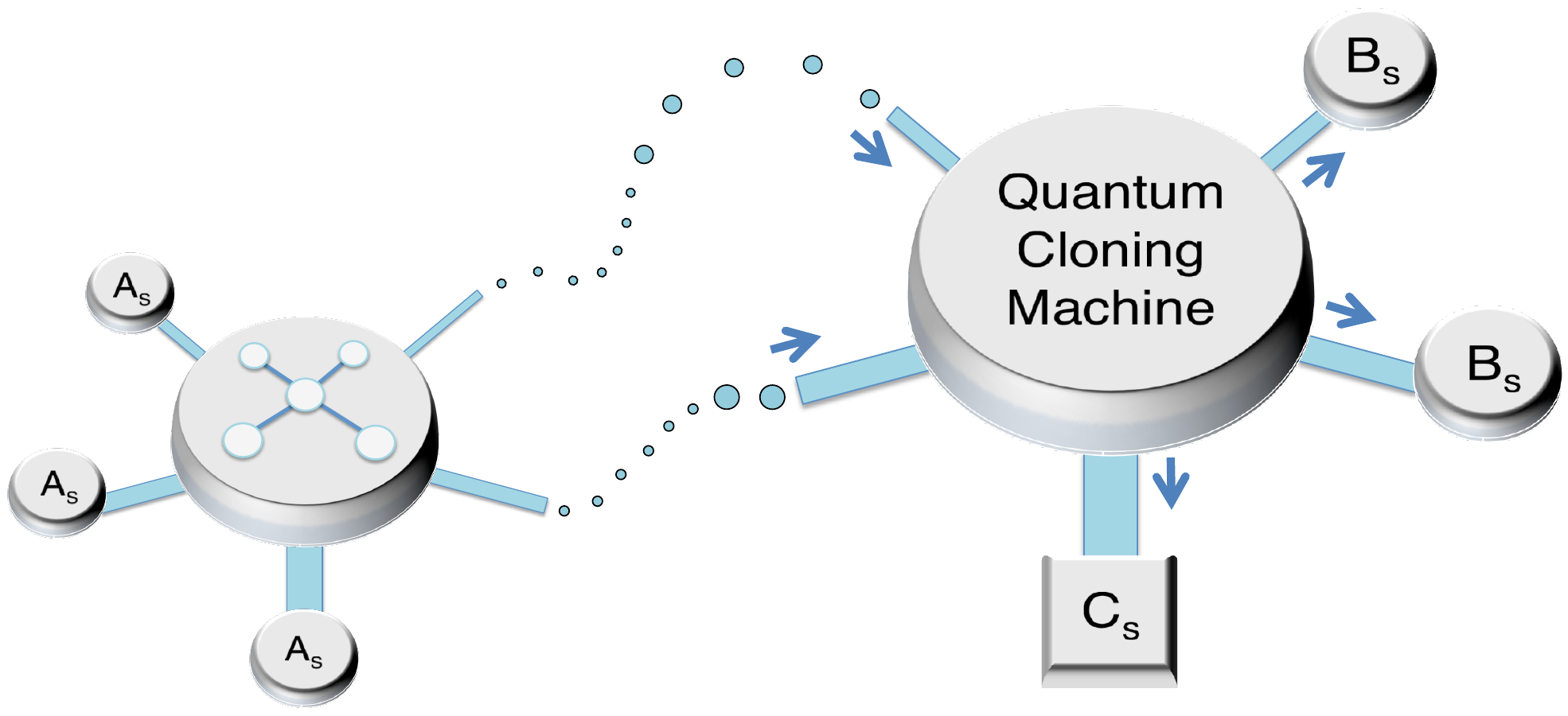}
\caption{(Color online). Sharing multipartite EPR steering with a quantum cloner. The party $C_{s}$ with an ancilla $C'_{s}$ (not shown) wants to share the multipartite EPR steerability of $\left|G_{2}\right\rangle$, between the groups $A_{s}$ and $B_{s}$, by using a quantum cloning machine. See Appendix C for detailed discussions.} \label{Cloning}
\end{figure}

\section{No-sharing of multipartite EPR steering}

A universal cloning machine can produce a clone of an unknown state with high fidelity \cite{Buzek96}. This result of quantum mechanics has significant implications in understanding quantum systems and profound applications in quantum information. Here we use it as an eavesdropping attack as used on the protocols of quantum cryptography \cite{Scarani05}.

Suppose $A_{s}$'s and $B_{s}$'s qudits are in a state $\left|G_{2}\right\rangle$ and, before receipt, $B_{s}$'s qudits are sent to a universal cloning machine \cite{Buzek96,Scarani05}. A third party, $C_{s}$, with an ancilla $C'_{s}$, receives some of the output qudits of the cloning machine (see~Fig.~\ref{Cloning}). We examine the mutual information between the results of measurements of $B_{s}$ and $C_{s}$ with those of $A_{s}$, where $A_{s}$, $B_{s}$ and $C_{s}$ implement the complementary measurements $\text{A}^{(S)}_{sm}$, $\text{B}^{(S)}_{sm}$ and $\text{C}^{(S)}_{sm}$, respectively, on their qudits in the Schmidt bases (see Appendix B.1 and B.2). Hence, we derive the following relationship between the mutual information of $B_{s}$ and $C_{s}$ with $A_{s}$
\begin{equation}
\sum_{m=1}^{2}I_{A_{sm}C_{sm}}+\sum_{m=1}^{2}I_{A_{sm}B_{sm}}\leq 2\log_{2}d,\label{iacfinal}
\end{equation}
for any multipartite graph states $\left|G_{2}\right\rangle$, including the simplest two-qudit graph state \cite{Chiu16}. See Appendix C for details of the derivation.

\begin{figure}[t]
\includegraphics[width=8.8cm,angle=0]{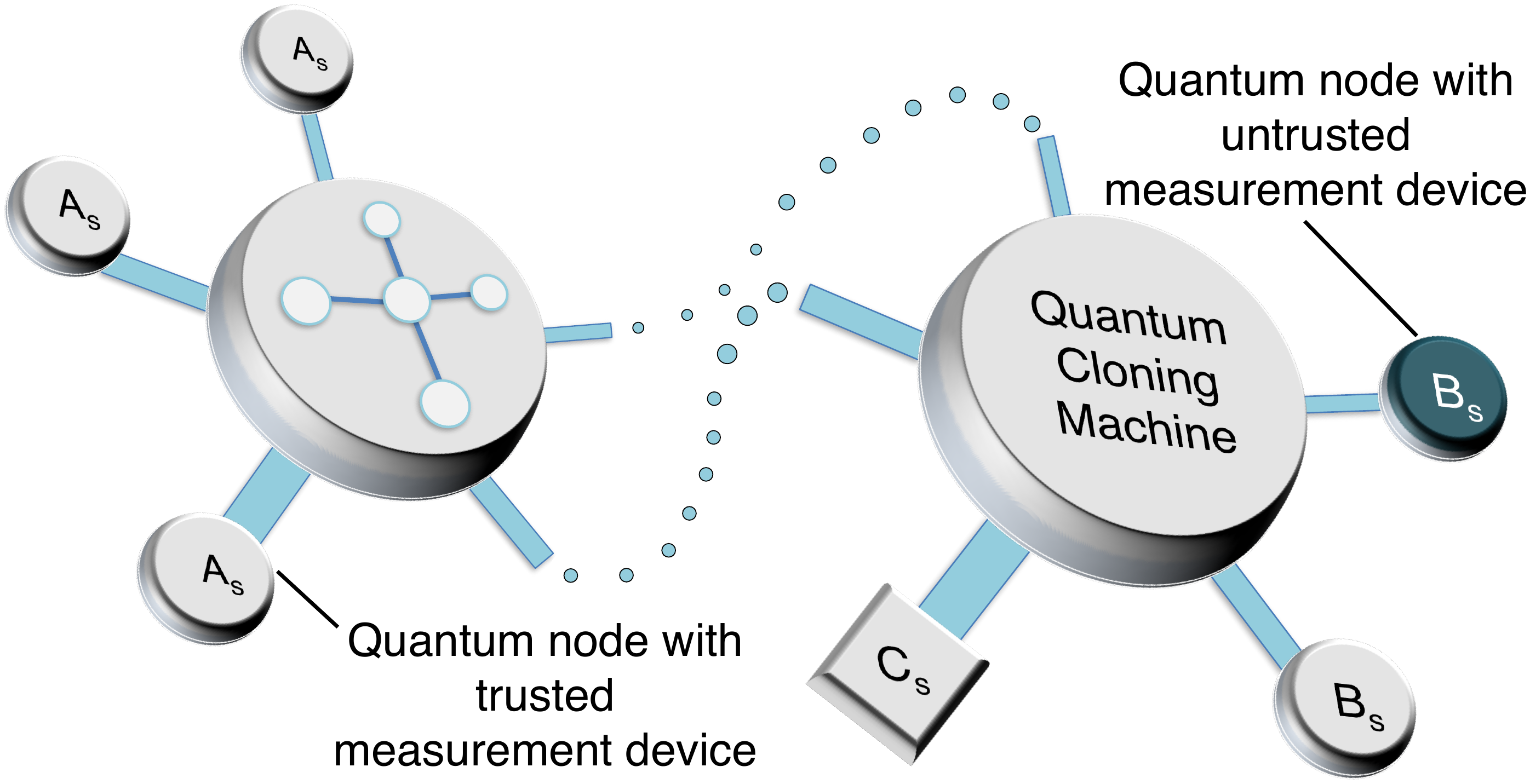}
\caption{(Color online). Quantum networks under both a quantum cloner attack and with untrusted measurements for quantum nodes. In addition to the cloning attack, the quantum-node holders may lose control over their measurement devices such that the system state may be in $\rho_{B_{s}}$~(\ref{rhobs}) in the worst case. For concreteness, it may happen in QSS where the untrusted parties in the group $B_{s}$ attempt to use their own measurements and the cloner to obtain knowledge of $A_{s}$'s key without collaboration with the rest of the trusted parties in $B_{s}$. The no-sharing of multipartite EPR steering can be used to secure QSS by excluding such possibility with the criteria (\ref{Icondition}) and (\ref{iacfinal}).} \label{QNeCU}
\end{figure}

The criterion (\ref{iacfinal}) reveals that, when the correlation between the qudits shared by $A_{s}$ and one of the two groups, say $B_{s}$, is identified as steering by Eq.~(\ref{Icondition}), the steering effect provides stronger correlations than the mutual dependence between $A_{s}$ and $C_{s}$ that cannot be replicated by a quantum cloning machine. To explain intuitively, criterion (8) concretely describes the total correlation that $A_{s}$ can share with $B_{s}$ and $C_{s}$ individually under cloning attacks. The importance of criterion (8) is further supported by the criterion (7) to confirm the steerability. In other words, multipartite EPR steering powers this type of non-classical mutual information between $A_{s}$ and $B_{s}$ that cannot be shared with the third party, $C_{s}$, by a universal quantum cloner. Hence the criteria (\ref{Icondition}, \ref{iacfinal}) can be used to rule out \emph{both} untrusted participants and cloning-based attacks for quantum networks, and thus can be exploited to secure a variety of quantum networking tasks (see Fig.~\ref{QNeCU}).

\section{Securing distributed quantum-information processing}

The no-sharing of multipartite steering has direct applications to quantum information protocols involving multiple participants. For example, for quantum computation, following the MBQC protocol \cite{Raussendorf01,Raussendorf03,Walther05,Chen07,Briegel09}, we assume that, $A_{s}$ and $B_{s}$ share a state $\left|G_{2}\right\rangle$, and that the inputs for a computation task are prepared by measurements on the qudits held by $A_{s}$. The outputs of the computation can then be obtained by performing local operations on $B_{s}$'s qudits according to $A_{s}$'s measurement results. The criterion (\ref{Icondition}) can quantitatively describe how the statistical dependence between $A_{s}$'s inputs and $B_{s}$'s outputs in terms of the mutual information $I_{A_{s}B_{s}}$ go beyond the ``cheating scenario'' using the states $\rho_{B_{s}}$. In particular, the criteria (\ref{Icondition}) and (\ref{iacfinal}) together imply that such dependence between inputs and outputs cannot be copied by the eavesdropper to deduce the computation result, which secures the quantum computation task. This concept and method can be extended and applied to BQC as well, to enable a client, who delegates a computation to a quantum server \cite{Broadbent09,Barz12}, to evaluate the uncharacterized facilities of the server and the security of the underlying quantum networks.

\begin{figure}[t]
\includegraphics[width=5.4cm,angle=0]{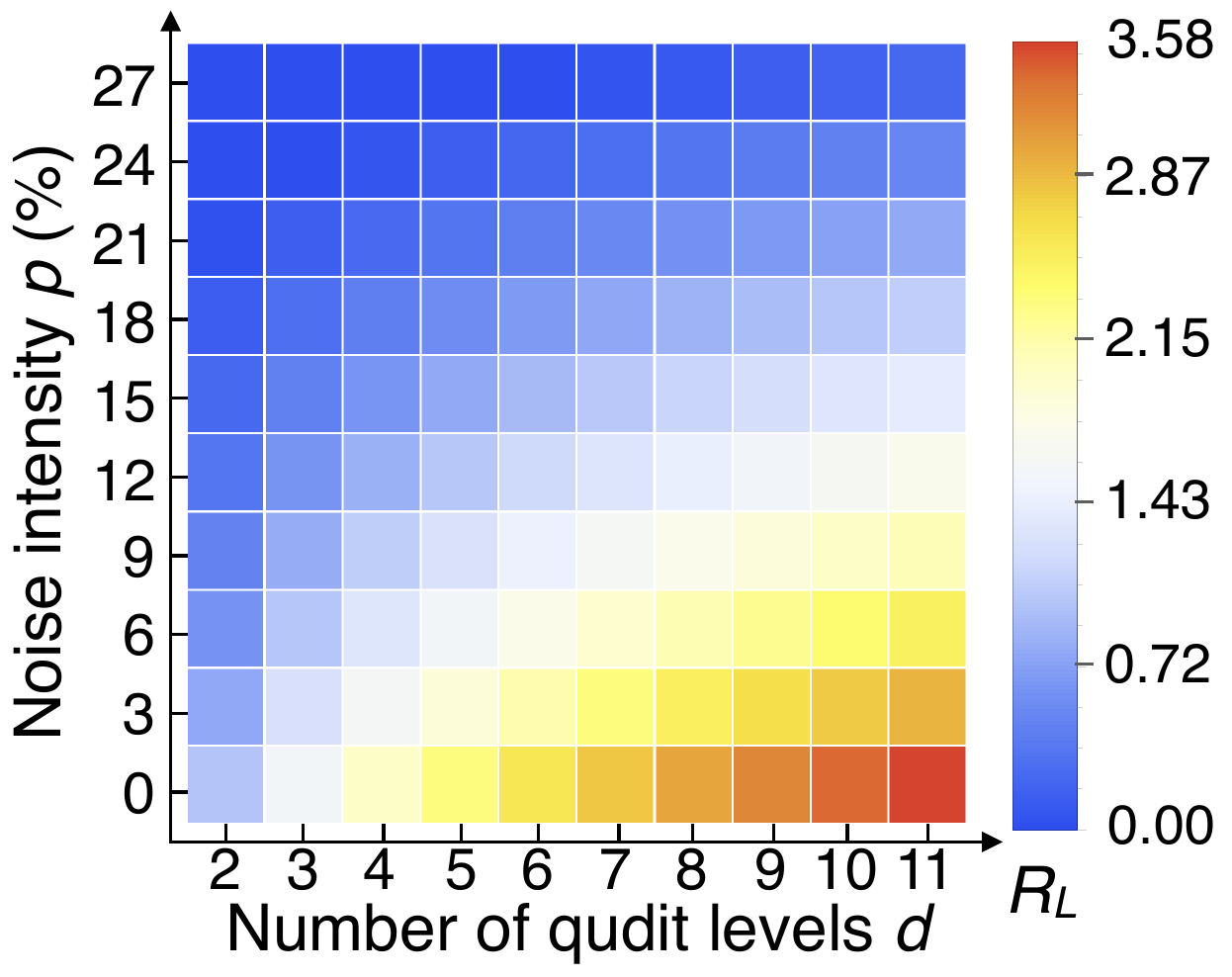}
\caption{(Color online). The lower bound of secret key rate $R_{L}$ derived from noisy graph states. As demonstrated, the reduction in $R_{L}$ by white noise (with intensity $p$) on $\left|G_{2}\right\rangle$ is independent of $N$. $R_{L}$ can be increased by increasing $d$. This trend is comparable to the increase in the certified multipartite steering with $d$, i.e., the noise tolerance of the criterion (\ref{Icondition}) (Appendix B.3). Here the measurements used for creating secret keys, $\text{A}_{sm}$ and $\text{B}_{sm}$, are extracted from Schmidt bases $\text{A}^{(S)}_{sm}$ and $\text{B}^{(S)}_{sm}$ \cite{Li10} and satisfy the uncertainty relation under the two POVMs \cite{Tomamichel11,Coles17} as introduced above, and each POVM is composed of two operator elements respectively. See Appendix B.1 for details. Note that, for any cases where $I_{A_{s}B_{s}}-\log_{2}d<0$, the corresponding key rates are set as zero.} \label{keyrate}
\end{figure}

In addition to quantum computation, the no-sharing restriction plays an important role in securing QSS \cite{Chen05,Hillery99,Markham08,Bell14a} and related applications, such as third-man quantum cryptography \cite{Chen05,Zukowski98} in the presence of untrusted participants. For example, $A_{s}$ is a dealer who sends a key to $B_{s}$. All the parties in $B_{s}$ are required by $A_{s}$ to collaborate to decode the key. If we assume that the dealer is trusted and $C_{s}$ is the eavesdropper who uses a quantum cloner to attack the quantum network between $A_{s}$ and $B_{s}$, the lower bound of the secret key rate for $A_{s}$ and $B_{s}$ can be determined by the Devetak-Winter formula \cite{DevetakWinter05}:
\begin{equation}
R\geq I_{A_{sm}B_{sm}}-\chi_{A_{sm}C_{sm}},\label{DW}
\end{equation}
where the Holevo quantity is defined by
\begin{equation}
\chi_{A_{sm}C_{sm}}\equiv S(\rho_{C_{s}C'_{s}})-\!\!\sum_{v_{Am}=0}^{d-1}P(v_{Am})S(\rho_{C_{s}C'_{s}|v_{Am}}).\nonumber
\end{equation}
Here $S(\rho_{C_{s}C'_{s}})$ is the von-Neumann entropy of the reduced state $\rho_{C_{s}C'_{s}}$, and $\rho_{C_{s}C'_{s}|v_{Am}}$ is the state conditioned on $A_{s}$'s result $v_{Am}$. Note that the role of $C_{s}$ can also be played by the untrusted parties in $B_{s}$, who lie about their measurements and use the quantum cloner to obtain maximal knowledge of the dealer's key without collaboration with the trusted parties, as described in the unconditional security proof for partially device-independent QSS protocols \cite{Kogias17} (see Fig.~\ref{QNeCU}).

Using a similar method as employed in Eq.~(\ref{iacfinal}) (see Appendix D for a complete derivation), we can arrive at the following lower bound for the secret key rate
\begin{equation}
R\geq I_{A_{s}B_{s}}-\log_{2}d.\label{R}
\end{equation}
The multipartite steerability for systems with arbitrary party number identified by the criteria (\ref{Icondition}, \ref{iacfinal}) guarantees that $B_{s}$ is trustworthy, and $A_{s}$ and $B_{s}$ can establish a secret key with a nonzero rate, by which the importance of multipartite steering to QSS for tripartite systems \cite{Xiang17} and even arbitrarily large systems can be appreciated.

Note that the magnitude of the mutual information $I_{A_{s}B_{s}}$, beyond the steering threshold $\log_{2}d$, determines the lower bound of the key rate:
\begin{equation}
R_{L}= I_{A_{s}B_{s}}-\log_{2}d. \nonumber
\end{equation}
As shown by Eq.~(\ref{iacfinal}), the attack of the quantum cloner can decrease the mutual information $I_{A_{s}B_{s}}$ and then reduce the lower bound of the key rate $R_{L}$. For the worst case, the error due to the quantum cloner even causes zero key rate, $R_{L}=0$. Such an error can be quantitively described by a value called the critical disturbance of the quantum cloner \cite{Scarani05}, $D_{c}$, and can be numerically determined, as shown in Appendix~D. For example, we have $D_{c}\approx 11\%$ for $d=2$. This exactly coincides with the existing result based on the best eavesdropping with a coherent attack for bipartite quantum key distribution \cite{Sheridan10}. Therefore, the attack with a cloning machine is optimal for this case. While it is not clear whether a quantum cloner is optimal for attacks on quantum networks with more than two participants, $N>2$, the optimal result for $N=2$ illustrates that there exists a deep relationship between the security of quantum communication and the no-cloning theorem. The criterion on the key rate (\ref{R}) based on the attack with a quantum cloner and the no-sharing of multipartite EPR steering could play a crucial role in ultimately securing generic quantum networking tasks.

In addition to the errors from the quantum cloner, any destructive influence on the steering reduces the lower bound of the key rate. See Fig.~\ref{keyrate} for concrete illustrations with noisy graph states. When transmitting graph states without suffering from any loss or interference, the key rate achieves the maximum: $R=\log_{2}d$, independent of the qudit number.

It is worth noting that testing the criterion (\ref{Icondition}) requires only two local measurement settings for each quantum node, which is naturally suitable for generic quantum-information protocols using graph states. For instance, the measurements $A_{s}$ and $B_{s}$ can be chosen to coincide with those required in a MBQC task \cite{Raussendorf01,Raussendorf03,Walther05,Chen07,Briegel09}.

\section{General quantum networking tasks}

In addition to distributed quantum information, all the graph-state-based networking tasks require the distribution of graph states. The criteria (\ref{Icondition},\ref{iacfinal}) enables a task verifier or trusted participants to actively examine whether the received states are capable of preventing eavesdroppers from learning \emph{any} task information with a quantum cloner, as demonstrated by the examples of MBQC, BQC and QSS. This secures both state sources and channels against cloning-based attacks.

\section{Conclusion and outlook}
We have developed a formalism to explore the role of multipartite EPR steerability of two-colarable graph states in securing distributed quantum-information tasks and showed that such high-order EPR steering cannot be shared by an eavesdropper using a universal quantum cloning machine, even in circumstances where a set of untrusted participants are involved. With a series of examples we illustrated how multipartite steering powers distributed quantum-information processing in a secure manner. We expect that our criteria secure the initialization of network nodes in the joint graph states for generic quantum networking tasks.

This conclusion motivates several questions for future work: apart from two-colorable graph states shown here, does this quantum characteristic exist in \emph{any} graph state? If this is the case, how do we confirm its existence in an experimentally efficient way? Moreover, in addition to multipartite steering, are the entropy-based criteria (\ref{Icondition}, \ref{iacfinal}) useful for verifying genuine multipartite EPR steerability of graph states? Finally, because the assumption of a trusted group is made for the steering criterion (\ref{Icondition}), how a verifier, such as the dealer in QSS, can perform a reliable and objective evaluation of which node can be identified as trusted or untrusted becomes critical for large-scale networking tasks. The error and imperfections in the creation and manipulation of graph states grow with the system size, which increases the participants' uncertainty about the created states and the total quantum network. This question poses an interesting and significant challenge for partially device-independent applications.

\acknowledgements

C.-M.L is partially supported by the Ministry of Science and Technology, Taiwan, under Grant Number MOST 104-2112-M-006-016-MY3 and MOST 107-2628-M-006-001-MY4. F.N. is supported in part by the: MURI Center for Dynamic Magneto-Optics via the Air Force Office of Scientific Research (AFOSR) (FA9550-14-1-0040), Army Research Office (ARO) (Grant No. Grant No. W911NF-18-1-0358), Asian Office of Aerospace Research and Development (AOARD) (Grant No. FA2386-18-1-4045), Japan Science and Technology Agency (JST) (Q-LEAP program, ImPACT program, and CREST Grant No. JPMJCR1676), Japan Society for the Promotion of Science (JSPS) (JSPS-RFBR Grant No. 17-52-50023, and JSPS-FWO Grant No. VS.059.18N), RIKEN-AIST Challenge Research Fund, and the John Templeton Foundation. N.L. acknowledges partial support from JST PRESTO, Grant No. JPMJPR18GC.

\appendix

\section{Two-colorable Graph states}

\subsection{State vectors}
A two-colorable graph has vertices that can be divided into two sets, where each set corresponds to a color such that adjacent vertices relate two different colors, such as star and chain graphs. See Fig.~\ref{GraphNet} and Fig.~\ref{graph}. For a given two-colorable graph $G(V,E)$ \cite{Hein04,Hein05} used for networking tasks, an edge, $(i,j)\in E$, corresponds to a unitary two-qudit transformation among the two qudits (vertices) $i$ and $j$, \begin{equation}
U_{(i,j)}\!\!=\!\!\sum_{v=0}^{d-1}\left|v\right\rangle_{ii}\!\left\langle v\right|\otimes (Z_{j})^{v},\label{U}
\end{equation}
where $\{\left|v\right\rangle_{i}|v\in\textbf{v}\}$, with $\textbf{v}=\{0,1,...,d-1\}$, is an orthonormal basis of the $i$th qudit and
\begin{equation}
Z_{j}=\sum_{v=0}^{d-1}\omega^{v}\left|v\right\rangle_{jj}\!\left\langle v\right|,
\end{equation}
with $\omega=\exp(i2\pi/d)$. The state vector of the target two-colorable colorable graph state can be obtained by applying $U_{(i,j)}$ to an initial state $\left|F_{0}\right\rangle=\bigotimes_{k=1}^{N}\hat{\text{F}}_{k}\left|0\right\rangle_{k}$ \cite{Zhou03} according to the edge set $E$:
\begin{equation}
\left|G_{2}\right\rangle=\prod_{(i,j)\in E}U_{(i,j)}\left|F_{0}\right\rangle,\label{G2}
\end{equation}
where $\hat{\text{F}}_{k}$ is the quantum Fourier transformation defined by $\hat{\text{F}}_{k}\left|v'\right\rangle_{k}=\sum_{v=0}^{d-1}\omega^{v'v}\left|v\right\rangle_{k}/\sqrt{d}$.

Regarding the topology of two-colorable graphs, it has been shown that genuine multipartite entanglement \cite{Li10} and genuine multipartite Einstein-Podolsky-Rosen (EPR) steering \cite{Li15} for states close to all two-colorable graph states $\left|G_{2}\right\rangle$ can be efficiently identified with two local measurement settings. This feature has also been used in deriving the entropic criterion for multipartite EPR steering (\ref{Icondition}) and the relationship (\ref{iacfinal}).

\subsection{The Schmidt form of $\left|G_{2}\right\rangle$}

\begin{figure}[t]
\includegraphics[width=8.4cm,angle=0]{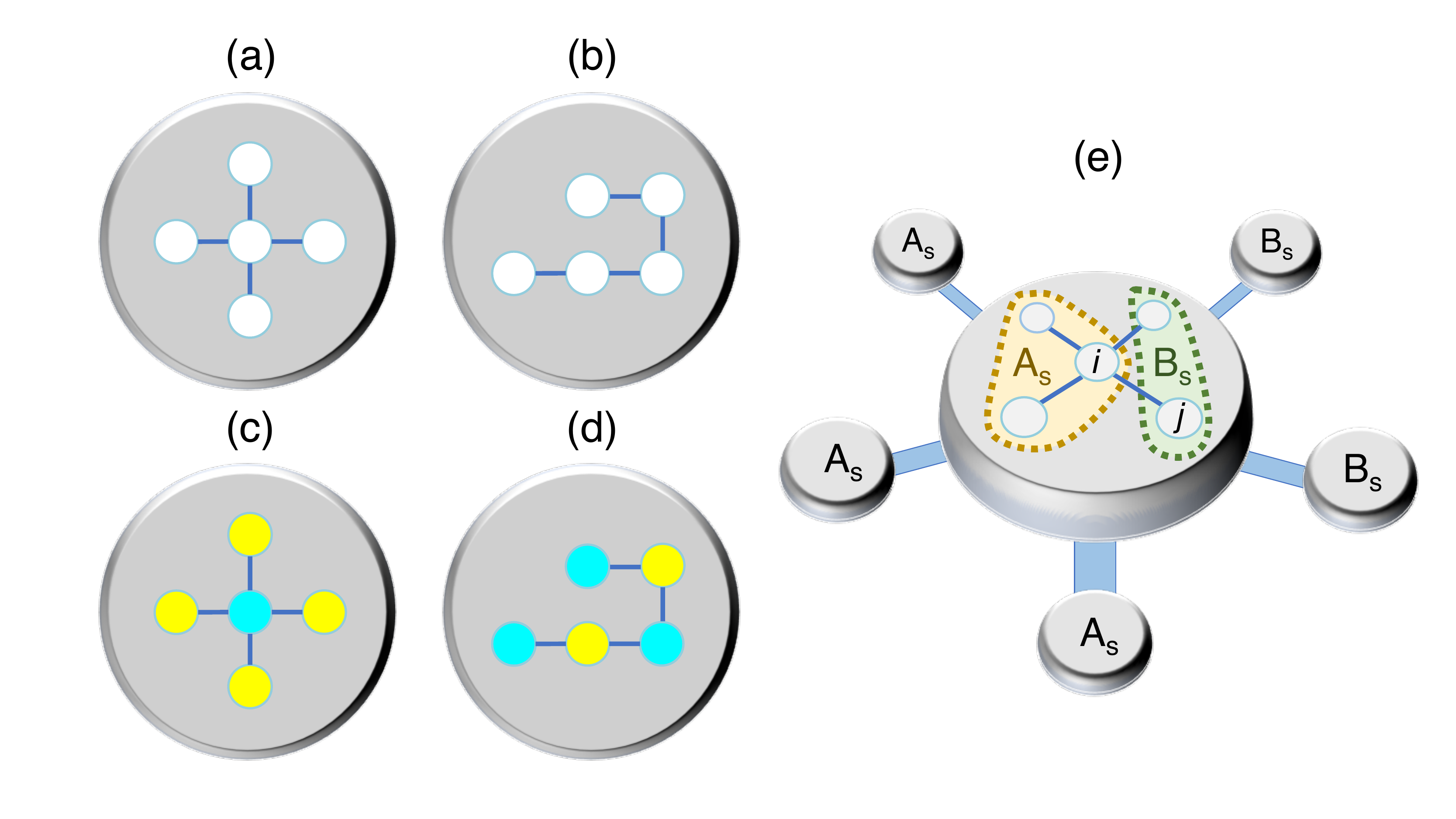}
\caption{(Color online) Two-colorable graph states. (a) Five-qudit star graph state. (b) Five-qudit chain graph state. The vertices of these graphs can be divided into two sets. The vertices of each set can be given a color such that adjacent vertices have different colors, as shown in (c) and (d). An edge, $(i,j)\in E$, corresponds to a unitary two-qudit transformation $U_{(i,j)}$~(\ref{U}) to cause non-classical correlation between qudits (e). The Schmidt form of a graph state, such as the star graph (e), with respect to $A_{s}$-subsystem (golden) and $B_{s}$-subsystem (green) can be shown by first choosing a qudit in $A_{s}$ (the $i$th qudit shown here) and then following the introduced procedure to find the state vector (\ref{G22}) for the final Schmidt decomposition~(\ref{Schmidt}).}\label{graph}
\end{figure}

For a bipartite splitting of $N$ quantum nodes of a $N$-qudit two-colorable graph state, one always can find a vertex in the $A_{s}$ subsystem, let us say the $i$th qudit ($i\in V_{A_{s}}$), with vertices in the $B_{s}$ subsystem forming edges $(i,j)\in E$, where $j\in V_{B_{s}}$. $V_{A_{s}}$ and $V_{B_{s}}$ denote the set of vertices of the subsystems $A_{s}$ and $B_{s}$, respectively, where $|V_{A_{s}}|+|V_{B_{s}}|=|V|=N$. See Fig.~\ref{graph}(e). When the $i$th qudit is represented in the basis $\{\left|v\right\rangle_{if}=\hat{\text{F}}_{i}\left|v\right\rangle_{i}|v\in\bold{v}\}$, the state vector of the graph state reads \cite{Li10}:
\begin{widetext}
\begin{equation}
\left|G_{2}\right\rangle=d^{-\frac{|N(i)|}{2}}\sum_{v_{1},...,v_{N};s_{ik}\doteq 0}[\left|v_{i}\right\rangle_{if}\bigotimes_{k\in V_{A_{s}}}(\left|v_{k}\right\rangle_{a}\left|v_{k}\right\rangle_{k})]\otimes[\bigotimes_{k\in V_{B_{s}}}(\left|v_{k}\right\rangle_{b}\left|v_{k}\right\rangle_{k})],\label{G22}
\end{equation}
\end{widetext}
where $s_{ik}=-v_{i}+\sum_{k\in N(i)}v_{k}$. The state vectors $\left|v_{k}\right\rangle_{a}$ and $\left|v_{k}\right\rangle_{b}$ are composed of qudits in $V_{A_{s}}$ and $V_{B_{s}}$, respectively, and accompanied by $\left|v_{k}\right\rangle_{k}$ for $k\in N(i)$, where $N(i)$ is the set of vertices that form edges with the $i$th vertex. For instance, we have $N(i)=4$ in Fig.~\ref{graph}(e).

Since the connection between $v_{i}$, $v_{j}$ and $v_{k}$ for $k\in N(i)$ is constrained by $s_{ik}=-v_{i}+\sum_{k\in{V_{A}}}v_{k}+\sum_{k\in{V_{B}}}v_{k}\doteq 0$, where $\doteq$ denotes equality modulo $d$, the state vector (\ref{G22}) can be explicitly represented as
\begin{equation}
\left|G_{2}\right\rangle=\frac{1}{\sqrt{d}}\sum_{v=0}^{d-1}\left|v\right\rangle_{A_{s}m}\otimes\left|v\right\rangle_{B_{s}m},\label{Schmidt}
\end{equation}
where
\begin{widetext}
\begin{equation}
\left|v\right\rangle_{A_{s}m}=d^{-\frac{|V_{A_{s}}|-1}{2}}\!\!\!\!\sum_{v_{1},...,v_{N};s_{ikv}\doteq 0}\!\!\!\!\left|v_{i}\right\rangle_{if}\bigotimes_{k\in V_{A_{s}}}(\left|v_{k}\right\rangle_{a}\left|v_{k}\right\rangle_{k}),\ \ \ \left|v\right\rangle_{B_{s}m}=d^{-\frac{|V_{B_{s}}|-1}{2}}\!\!\!\!\sum_{v_{1},...,v_{N};s_{k}\doteq v}\bigotimes_{k\in V_{B_{s}}}(\left|v_{k}\right\rangle_{b}\left|v_{k}\right\rangle_{k}),\label{SchmidtVectors}
\end{equation}
\end{widetext}
$s_{ikv}=-v_{i}+\sum_{k\in{V_{A_{s}}}}v_{k}+v$, and $s_{k}=\sum_{k\in{V_{B_{s}}}}v_{k}$. The state vectors $\left|v\right\rangle_{A_{s}m}$'s and $\left|v\right\rangle_{B_{s}m}$'s constitute two orthonormal bases $\text{A}^{(S)}_{sm}=\{\left|v\right\rangle_{A_{s}m}\}$ and $\text{B}^{(S)}_{sm}=\{\left|v\right\rangle_{B_{s}m}\}$. Hence the above representation is the Schmidt form of $\left|G_{2}\right\rangle$, as shown in Eq.~(1) in the main text. Note that the subscript $m=1,2$ reminds that the state $\left|G_{2}\right\rangle$ can be represented in two different Schmidt bases $(\text{A}^{(S)}_{sm},\text{B}^{(S)}_{sm})$, which are complementary to each other.

\section{Steering Criterion}

\subsection{Measurements in the steering criterion}

The design of the measurements required to implement the steering criterion (\ref{Icondition}) is based on the characteristics of the two-colorable graph states represented in the Schmidt bases~(\ref{Schmidt}). For all two-colorable graph states, only the minimum two local measurement settings $(\text{A}_{s1},\text{B}_{s1})$ and $(\text{A}_{s2},\text{B}_{s2})$ are sufficient to verify multipartite EPR steering.

The measurement outcomes $\{a_{s,1},b_{s,1}\}$ and $\{a_{s,2},b_{s,2}\}$ obtained from the measurements $(\text{A}_{s1},\text{B}_{s1})$ and $(\text{A}_{s2},\text{B}_{s2})$, respectively, are used to determine the mutual information $I_{A_{s}B_{s}}$ for the steering criterion. Here $\text{A}_{sm}$ and $\text{B}_{sm}$ are general as POVMs. The POVM operator elements for each POVM depend on the type of the bipartite splitting and the characteristics of the target graph state, and each POVM operator element can be locally measured on individual quantum nodes. These operators are extracted from the basis vectors (\ref{SchmidtVectors}) of the Schmidt bases $\text{A}^{(S)}_{sm}$ and $\text{B}^{(S)}_{sm}$ such that they are commutative with the measurement operators in the Schmidt bases. Moreover, when measuring with the two POVMs, $\text{B}_{s1}$ and $\text{B}_{s2}$, which are complementary bases, the entropic uncertainty relation (\ref{uncertain}) always holds for the quantum states $\rho_{\lambda_{\alpha}}$ [see Eq.~(\ref{rhobs})].

To elaborate, here we give a concrete example of a 3-qubit star-graph state, where qubit 1 in $A_{s}$ is connected with qubit 2 and qubit 3 in $B_{s}$. According to (\ref{Schmidt},\ref{SchmidtVectors}), its state vector can be expressed as
\begin{equation}
\left|G_{star}\right\rangle=\frac{1}{\sqrt{2}}\sum_{v=0}^{1}\left|v\right\rangle_{A_{s}m}\otimes\left|v\right\rangle_{B_{s}m},\nonumber
\end{equation}
where
\begin{eqnarray}
&&\text{A}^{(S)}_{s1}=\{\left|0\right\rangle_{A_{s}1}=\left|0\right\rangle_{1},\left|1\right\rangle_{A_{s}1}=\left|1\right\rangle_{1}\},\nonumber\\
&&\text{B}^{(S)}_{s1}=\{\left|0\right\rangle_{B_{s}1}=\left|++\right\rangle_{23},\left|1\right\rangle_{B_{s}1}=\left|--\right\rangle_{23}\},\nonumber\\
&&\text{A}^{(S)}_{s2}=\{\left|0\right\rangle_{A_{s}2}=\left|+\right\rangle_{1},\left|1\right\rangle_{A_{s}2}=\left|-\right\rangle_{1}\},\nonumber\\
&&\text{B}^{(S)}_{s2}=\{\left|0\right\rangle_{B_{s}2}=\frac{1}{\sqrt{2}}(\left|00\right\rangle_{23}+\left|11\right\rangle_{23}),\nonumber\\
&&\ \ \ \ \ \ \ \ \ \ \ \ \left|1\right\rangle_{B_{s}2}=\frac{1}{\sqrt{2}}(\left|01\right\rangle_{23}+\left|10\right\rangle_{23})\},\nonumber
\end{eqnarray}
and $\left|\pm\right\rangle=(\left|0\right\rangle\pm\left|1\right\rangle)/\sqrt{2}$. By examining the Schmidt bases, $\text{A}^{(S)}_{s1}$, $\text{B}^{(S)}_{s1}$, $\text{A}^{(S)}_{s2}$, and $\text{B}^{(S)}_{s2}$, we then obtain the following POVMs in the locally measurable bases:
\begin{eqnarray}
&&\text{A}_{s1}= \{\left|0\right\rangle_{11}\!\left\langle0\right|,\left|1\right\rangle_{11}\!\left\langle1\right|\},\nonumber\\
&&\text{B}_{s1}=\{\left|++\right\rangle_{2323}\!\left\langle++\right|+\left|+-\right\rangle_{2323}\!\left\langle+-\right|,\nonumber\\
&& \ \ \ \ \ \ \ \ \ \ \left|-+\right\rangle_{2323}\!\left\langle-+\right|+\left|--\right\rangle_{2323}\!\left\langle--\right|\},\nonumber\\
&&\text{A}_{s2}= \{\left|+\right\rangle_{11}\!\left\langle+\right|,\left|-\right\rangle_{11}\!\left\langle-\right|\},\nonumber\\
&&\text{B}_{s2}=\{\left|00\right\rangle_{2323}\!\left\langle00\right|+\left|11\right\rangle_{2323}\!\left\langle11\right|,\nonumber\\
&& \ \ \ \ \ \ \ \ \ \ \left|01\right\rangle_{2323}\!\left\langle01\right|+\left|10\right\rangle_{2323}\!\left\langle10\right|\}.\nonumber
\end{eqnarray}
It is clear that these POVM operator elements are commutative with the Schmidt bases such that Eq.~(\ref{MI}) holds, and the trusted and untrusted roles of $A_{s}$ and $B_{s}$ can be exchanged. Following the same procedure, our method can be easily extended to the graph states with arbitrary $N$ and $d$.

\subsection{Criterion in the Schmidt bases}

In addition to the steering criterion in the form (\ref{Icondition}) under the measurement settings $(\text{A}_{sm},\text{B}_{sm})$, the steering condition can also be concretely represented in terms of the mutual information under the Schmidt bases $(\text{A}^{(S)}_{sm},\text{B}^{(S)}_{sm})$ as follow,
\begin{equation}
I^{(S)}_{A_{s}B_{s}}\equiv\sum_{m=1}^{2}I^{(S)}_{A_{sm}B_{sm}}> \sum_{m=1}^{2}I^{(S)}_{\lambda_{\alpha}B_{sm}},\label{Is2}
\end{equation}
where $I^{(S)}_{A_{sm}B_{sm}}$ and $I^{(S)}_{\lambda_{\alpha}B_{sm}}$ are the mutual information between their results derived from the measurements $\text{A}^{(S)}_{sm}$ and $\text{B}^{(S)}_{sm}$ in the Schmidt bases. The measurement outcomes of $A^{(S)}_{s}$ and $B^{(S)}_{s}$: $\{a^{(S)}_{s,1},b^{(S)}_{s,1}\}$ and $\{a^{(S)}_{s,2},b^{(S)}_{s,2}\}$, are then obtained from the measurements which corresponds to $(\text{A}^{(S)}_{s1},\text{B}^{(S)}_{s1})$ and $(\text{A}^{(S)}_{s2},\text{B}^{(S)}_{s2})$, respectively. The entropy of $B_{s}$'s outcomes and the entropy conditioned on $A_{s}$'s results are therefore derived from these measurement results. In addition, since the unknown quantum state $\rho_{\lambda_{\alpha}}$ satisfies the entropic uncertainty relation under the two complementary bases, $\text{B}^{(S)}_{s1}$ and $\text{B}^{(S)}_{s2}$ \cite{Tomamichel11,Coles17}, the steering criterion~(\ref{Is2}) in the Schmidt bases becomes
\begin{equation}
I^{(S)}_{A_{s}B_{s}} > \log_{2}d.\label{Iconditionsch}
\end{equation}
For any two-colorable graph states, as illustrated in Eq.~(\ref{MI}), quantum measurements on the qudits can show that
\begin{equation}
I^{(S)}_{A_{s}B_{s}}=2\log_{2}d.\nonumber
\end{equation}
Therefore, their dependence is stronger than the correlation between $B_{s}$ and $\rho_{B_{s}}$.

\subsection{Noise tolerance}

\begin{figure}[t]
\includegraphics[width=7.4cm,angle=0]{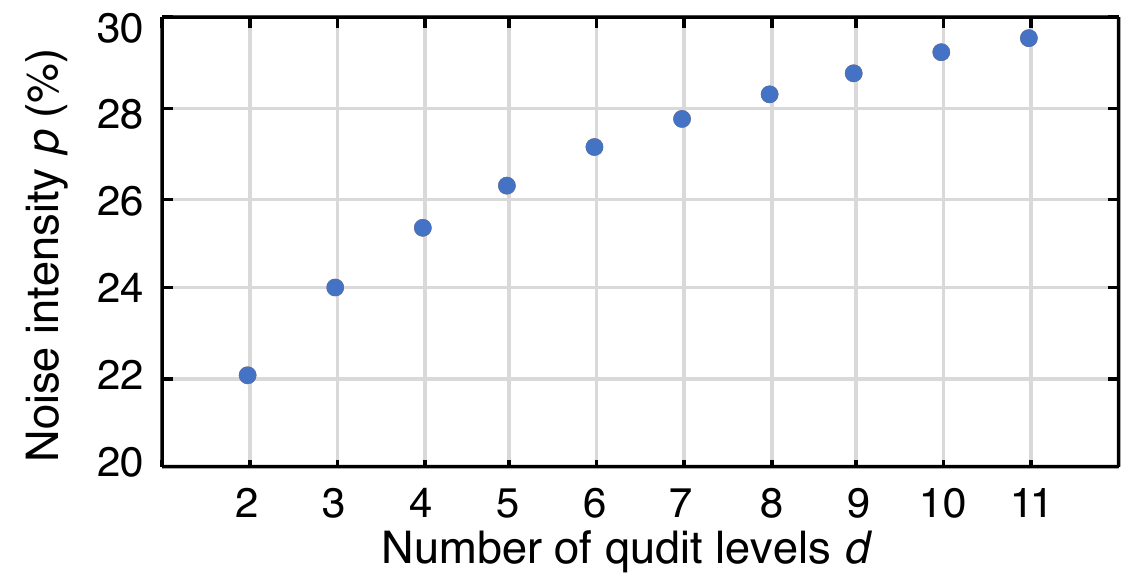}
\caption{(Color online). Noise tolerance of the steering criterion: $I_{A_{s}B_{s}} > \log_{2}d$, for arbitrarily two-colorable graph states.} \label{tolerance}
\end{figure}

To examine the steering criterion from the viewpoint of robustness against noise, we consider the minimum amount of uncolored noise added to $\left|G_{2}\right\rangle$ such that the noisy state cannot be identified by the steering criterion (\ref{Icondition}), i.e.,
\begin{equation}
I_{A_{s}B_{s}} = \log_{2}d.\nonumber
\end{equation}
Suppose that, in the presence of white noise, the pure state $\left|G_{2}\right\rangle$ becomes
\begin{equation}
\rho_{G_{2}}(p)=\frac{p}{d^{N}}\hat{\text{I}}+(1-p)\left|G_{2}\right\rangle\!\!\left\langle G_{2}\right|,\nonumber
\end{equation}
where $p$ is the intensity of uncolored noise. The noise tolerance of criterion (\ref{Icondition}) is quantified by the noise threshold $p_{\text{noise}}$, such that if $p<p_{\text{noise}}$ then
\begin{equation}
I_{A_{s}B_{s}}(\rho_{G_{2}}(p))> \log_{2}d.\nonumber
\end{equation}
See Fig.~\ref{tolerance}. Here the $\text{A}_{sm}$ and $\text{B}_{sm}$ are extracted from Schmidt bases (\ref{SchmidtVectors}) and satisfy the uncertainty relation under the two POVMs \cite{Tomamichel11,Coles17} as introduced above, and each POVM is composed of two operator elements. It is worth noting that this certification is independent of the node number $N$, and becomes more robust against noise as the dimension of the quantum nodes (qudits) increases.

\section{No-sharing of multipartite Einstein-Podolsky-Rosen steering}

In what follows, we will prove Eq.~(\ref{iacfinal}) introduced in the main text for no-sharing of multipartite EPR steering. This criterion describes the relationship between the mutual information of $B_{s}$ and $C_{s}$ with $A_{s}$. First, we consider $\left|G_{2}\right\rangle$ to be an input of a quantum cloner. See Fig.~2. After cloning \cite{Buzek96,Scarani05}, the output state of the total system becomes
\begin{equation}
\left|\Psi\right\rangle_{A_{s}B_{s}C_{s}C'_{s}}=\sum_{j,k=0}^{d-1}\sqrt{\gamma_{jk}}\left|\Psi_{jk}\right\rangle_{A_{s}B_{s}}\left|\Psi_{j,d-k}\right\rangle_{C_{s}C'_{s}},
\end{equation}
where
\begin{equation}
\left|\Psi_{jk}\right\rangle_{n'n}=\frac{1}{\sqrt{d}}\sum_{v=0}^{d-1}\omega^{vk} \left| v\right\rangle_{n'1}\left| v+_{\scriptscriptstyle{d}}j\right\rangle_{n1},
\end{equation}
with $\left|\Psi_{00}\right\rangle_{n'n}=\left|G_{2}\right\rangle$ for $(n',n)=(A_{s},B_{s}),(C_{s},C'_{s})$, and $+_{\scriptscriptstyle{d}}$ denotes addition modulo $d$. The qudits of $A_{s}$ and $B_{s}$ are in the reduced state
\begin{equation}
\rho_{A_{s}B_{s}}=\sum_{j,k=0}^{d-1}\gamma_{jk}\left|\Psi_{jk}\right\rangle_{A_{s}B_{s} A_{s}B_{s}}\!\left\langle \Psi_{jk}\right|,
\end{equation}
and $C_{s}$'s qudits with the ancilla $C'_{s}$ have the reduced state
\begin{equation}
\rho_{C_{s}C'_{s}}=\sum_{j,k=0}^{d-1}\gamma_{jk}\left|\Psi_{j,d-k+1}\right\rangle_{C_{s}C'_{s}C_{s}C'_{s}}\!\left\langle \Psi_{j,d-k+1}\right|.
\end{equation}

The mutual information between $A_{s}$ and $B_{s}$ of the reduced state $\rho_{A_{s}B_{s}}$ is
\begin{equation}
I_{A_{sm}B_{sm}}^{(S)}=\log_{2}d+\sum_{t=0}^{d-1}q_{m}^{t}\log_{2}q_{m}^{t},\label{Iqt}
\end{equation}
where $q_{1}^{t}=\sum_{k=0}^{d-1}\gamma_{tk}$ and $q_{2}^{t}=\sum_{j=0}^{d-1}\gamma_{j,d-t+1}$. The variables $q_{m}^{t}$ denote the probabilities of observing $v_{Bm}-v_{Am}=t$ or $v_{Bm}-v_{Am}=t-d$, for $t\in\textbf{v}$ \cite{Sheridan10}. Their sum is then
\begin{equation}
\sum_{m=1}^{2}I_{A_{sm}B_{sm}}^{(S)}=2\log_{2}d-\sum_{m=1}^{2}H(q_{m}^{t}).\label{iabfinal}
\end{equation}

To determine the mutual information, $I_{A_{sm}C_{sm}}^{(S)}$, between the results of measurements $\text{A}^{(S)}_{sm}$ and $\text{C}^{(S)}_{sm}$, we first consider the mutual dependence between $v_{Am}$ and the results derived from measurements on $C_{s}$'s qudits and ancilla $C'_{s}$ by their mutual information $I_{A_{sm}(C_{sm}C'_{sm})}^{(S)}$. It is clear that
\begin{equation}
I_{A_{sm}C_{sm}}^{(S)}\leq I_{A_{m}(C_{sm}C'_{sm})}^{(S)}.\label{iacacc}
\end{equation}
In addition, $I_{A_{sm}(C_{sm}C'_{sm})}^{(S)}$ is constrained by the Holevo bound by
\begin{equation}
 I_{A_{sm}(C_{sm}C'_{sm})}^{(S)}\leq \chi_{A_{sm}C_{sm}},\label{Hbound}
\end{equation}
where the Holevo quantity is
\begin{equation}
\chi_{A_{sm}C_{sm}}=S(\rho_{C_{s}C'_{s}})-\!\!\sum_{v_{Am}=0}^{d-1}P(v_{Am})S(\rho_{C_{s}C'_{s}|v_{Am}}).\nonumber
\end{equation}
Here, the von-Neumann entropy of the reduced state $\rho_{C_{s}C'_{s}}$ is defined by
\begin{equation}
S(\rho_{C_{s}C'_{s}})=-\sum_{j,k=0}^{d-1}\gamma_{jk}\log_{2}\gamma_{jk}\equiv H(\gamma).\label{s1}
\end{equation}
$\rho_{C_{s}C'_{s}|v_{Am}}$ is the state conditioned on $A_{s}$'s result $v_{Am}$, and the von-Neumann entropy of which is
\begin{equation}
S(\rho_{C_{s}C'_{s}|v_{Am}})=-\sum_{t=0}^{d-1}q_{m}^{t}\log_{2}q_{m}^{t}\equiv H(q_{m}^{t}).\label{s2}
\end{equation}

In order to derive the upper bound of $ I_{A_{sm}(C_{sm}C'_{sm})}^{(S)}$ by examining the difference between $S(\rho_{C_{s}C'_{s}})$ and $\sum_{v_{Am}=0}^{d-1}P(v_{Am})S(\rho_{C_{s}C'_{s}|v_{Am}})$, we substitute $\gamma_{j,d-k}=g(j,k)q_{1}^{j}$ into $q_{2}^{t}=\sum_{j=0}^{d-1}\gamma_{j,d-t}$, where $\sum_{k=0}^{d-1}g(j,k)=1$, and then obtain $q_{2}^{t}=\sum_{k=0}^{d-1}g(t,k)q_{1}^{k}$. For each $t$ all $g(t,k)=q_{2}^{t}$ shows the maximum of the difference. Then we have
\begin{equation}
 H(\gamma)=H(q_{1}^{t})+\sum_{t}q_{1}^{t}H(f(t))=H(q_{1}^{t})+H(q_{2}^{t}).\label{hq12}
\end{equation}
With Eqs.~(\ref{iacacc})-(\ref{hq12}), the upper bound of the mutual information $I_{A_{sm}C_{sm}}^{(S)}$ is then shown as
\begin{equation}
I_{A_{sm}C_{sm}}^{(S)}\leq H(q_{1}^{t})+H(q_{2}^{t})-H(q_{m}^{t}),\nonumber
\end{equation}
which implies that
\begin{equation}
\sum_{m=1}^{2}I_{A_{sm}C_{sm}}^{(S)}\leq \sum_{m=1}^{2}H(q_{m}^{t}).\label{iactemp}
\end{equation}

Combining Eq.~(\ref{iabfinal}) with Eq.~(\ref{iactemp}), we obtain
\begin{equation}
\sum_{m=1}^{2}I_{A_{sm}C_{sm}}^{(S)}+\sum_{m=1}^{2}I_{A_{sm}B_{sm}}^{(S)}\leq 2\log_{2}d.\label{iacfinalsch}
\end{equation}
For the simple bipartite case $N=2$, the above relation recovers the criterion used by Chiu \textit{et al.} \cite{Chiu16} to show no-cloning of EPR steering. For general $N\geq 3$, since the measurement operators in the Schmidt bases and those in the locally measurable bases $\text{A}_{sm}$, $\text{B}_{sm}$ and $\text{C}_{sm}$ commute with each other, we have the relations
\begin{equation}
\sum_{m=1}^{2}I_{A_{sm}C_{sm}}^{(S)}=\sum_{m=1}^{2}I_{A_{sm}C_{sm}}\label{isac}
\end{equation}
and
\begin{equation}
\sum_{m=1}^{2}I_{A_{sm}B_{sm}}^{(S)}=\sum_{m=1}^{2}I_{A_{sm}B_{sm}}.\label{isab}
\end{equation}
Thus, through Eqs.~(\ref{iacfinalsch}-\ref{isab}), we arrive at Eq.~(\ref{iacfinal}):
\begin{equation}
\sum_{m=1}^{2}I_{A_{sm}C_{sm}}+\sum_{m=1}^{2}I_{A_{sm}B_{sm}}\leq 2\log_{2}d,\nonumber
\end{equation}
As the correlation between the qudits shared by $A_{s}$ and $B_{s}$, is identified as multipartite steering by Eq.~(\ref{Icondition}), the mutual dependence between $A_{s}$ and $C_{s}$ then cannot show the steering effect.

\section{Lower bound of the secret key rate for quantum secret sharing blue and the critical disturbance of the quantum cloner}
To determine the lower bond of the secret key rate for QSS, as described by Eq. (\ref{DW}) \cite{DevetakWinter05}, we consider the following quantity
\begin{equation}
I_{A_{sm}B_{sm}}-\max \chi_{A_{sm}C_{sm}}.\nonumber
\end{equation}
From Eqs. (\ref{s1}-\ref{hq12},\ref{isab},\ref{isac}) and
\begin{equation}
I_{A_{sm}B_{sm}}=\log_{2}d-H(q_{m}^{t}),
\end{equation}
we get $\max \chi_{A_{sm}C_{sm}}= 2\log_{2}d-I_{A_{s}B_{s}}-H(q_{m}^{t})$. Therefore we obtain the following lower bound of the secret rate
\begin{equation}
R_{L}=I_{A_{s}B_{s}}-\log_{2}d,\label{rl}
\end{equation}
as shown in Eq.~(\ref{R}). The multipartite steerability identified by the criterion (\ref{Icondition}) then enables $A_{s}$ and $B_{s}$ to collaboratively generate a secret key with a nonzero rate. Combined with the noise tolerance obtained above, we can thus find the lower bound of secret key rate $R_L$ derived from noisy graph states, as illustrated in Fig. 4 in the main text.

Equation~(\ref{rl}) can be used to evaluate the critical disturbance of the quantum cloner that makes $R_L=0$. We first note that the variables $q_{m}^{t}$ for $t\neq 0$ quantitatively describe the errors introduced by the cloner. See the explanation for the $q_{m}^{t}$ in Eq.~(\ref{Iqt}). Then, suppose that the quantum cloner is phase-covariant \cite{Scarani05} which copies equally well the states of both bases, we have $H(q_{1}^{t})=H(q_{2}^{t})=H(D)$, where
\begin{equation}
H(D)=-(1-D)\log_{2}(1-D)-D\log_{2}\frac{D}{d-1},
\end{equation}
and $D=1-q_{m}^{0}$ is the disturbance due to the attack of quantum cloner. With $H(D)$, the critical disturbance, $D_{c}$, for $R_L=0$, can be derived by solving the equation
\begin{equation}
(1-D_{c})\log_{2}(1-D_{c})+D\log_{2}\frac{D_{c}}{d-1}=-\frac{1}{2}\log_{2}d.
\end{equation}
For example, the critical disturbance is $D_{c}\approx 11.00\%$ for $d=2$, and $D_{c}\approx 15.95\%$ for $d=3$.


\begin{references}

\bibitem{Schrodinger35} E. Schr\"odinger, Discussion of probability relations between separated systems, Math. Proc. Cambridge Philos. Soc. \textbf{31}, 555 (1935).

\bibitem{EPR35} A. Einstein, B. Podolsky, and N. Rosen, Can quantum-mechanical description of physical reality be considered complete?, Phys. Rev. \textbf{47}, 777 (1935).

\bibitem{Wiseman07} H. M. Wiseman, S. J. Jones, and A. C. Doherty, Steering, Entanglement, Nonlocality, and the Einstein-Podolsky-Rosen Paradox, Phys. Rev. Lett. \textbf{98}, 140402 (2007).

\bibitem{Branciard12} C. Branciard, E. G. Cavalcanti, S. P. Walborn, V. Scarani, and H. M. Wiseman, One-sided device-independent quantum key distribution: Security, feasibility, and the connection with steering, Phys. Rev. A \textbf{85}, 010301(R) (2012).

\bibitem{Kimble08} H. J. Kimble, The quantum internet, Nature (London) \textbf{453} 1023 (2008).

\bibitem{Duan10} L.-M. Duan and C. Monroe, \emph{Colloquium}: Quantum networks with trapped ions, Rev. Mod. Phys. \textbf{82}, 1209 (2010).

\bibitem{Northup14} T. E. Northup and R. Blatt, Quantum information transfer using photons, Nat. Photonics \textbf{8}, 356 (2014).

\bibitem{Reiserer15} A. Reiserer and G. Rempe, Cavity-based quantum networks with single atoms and optical photons, Rev. Mod. Phys. \textbf{87}, 1379 (2015).

\bibitem{Hein04} M. Hein, J. Eisert, and H. J. Briegel, Multiparty entanglement in graph states, Phys. Rev. A \textbf{69}, 062311 (2004).

\bibitem{Hein05} M. Hein, W. D\"ur, J. Eisert, R. Raussendorf, M. V. den Nest, and H. J. Briegel, Entanglement in Graph States and its Applications, arXiv:quant-ph/0602096.

\bibitem{Hillery99} M. Hillery, V. Bu\v{z}ek, and A. Berthiaume, Quantum secret sharing, Phys. Rev. A \textbf{59}, 1829 (1999).

\bibitem{Chen05} Y.-A. Chen, A.-N. Zhang, Z. Zhao, X.-Q. Zhou, C.-Y. Lu, C.-Z. Peng, T. Yang, and J.-W. Pan, Experimental quantum secret sharing and third-man quantum cryptography, Phys. Rev. Lett. \textbf{95}, 200502 (2005).

\bibitem{Markham08} D. Markham and B. C. Sanders, Graph States for Quantum Secret Sharing, Phys. Rev. A \textbf{78}, 042309 (2008).

\bibitem{Bell14a} B. A. Bell, D. Markham, D. A. Herrera-Mart\'i, A. Marin, W. J. Wadsworth, J. G. Rarity, and M. S. Tame, Experimental demonstration of graph-state quantum secret sharing, Nat. Commun. \textbf{5}, 5480 (2014).

\bibitem{Raussendorf01} R. Raussendorf and H. J. Briegel, A One-Way Quantum Computer, Phys. Rev. Lett. \textbf{86}, 5188--5191 (2001).

\bibitem{Raussendorf03} R. Raussendorf, D. E. Browne, and H. J. Briegel, Measurement-based quantum computation with cluster states, Phys. Rev. A \textbf{68}, 022312 (2003).

\bibitem{Walther05} P. Walther, K. J. Resch, T. Rudolph, E. Schenck, H. Weinfurter, V. Vedral, M. Aspelmeyer, and A. Zeilinger, Experimental one-way quantum computing, Nature (London) \textbf{434}, 169 (2005).

\bibitem{Chen07} K. Chen, C.-M. Li, Q. Zhang, Y.-A. Chen, A. Goebel, S. Chen, A. Mair, and J.-W. Pan, Experimental Realization of One-Way Quantum Computing with Two-Photon Four-Qubit Cluster States, Phys. Rev. Lett. \textbf{99}, 120503 (2007).

\bibitem{Briegel09} H. J. Briegel, D. E. Browne, W. D\"ur, R. Raussendorf, and M. Van den Nest, Measurement-based quantum computation, Nat. Phys. \textbf{5}, 19 (2009).

\bibitem{Schlingemann01} D. Schlingemann and R. F. Werner, Quantum error-correcting codes associated with graphs, Phys. Rev. A \textbf{65}, 012308 (2001).

\bibitem{Looi2008} S. Y. Looi, L. Yu, V. Gheorghiu, and R. B. Griffiths, Quantum-error-correcting codes using qudit graph states, Phys. Rev. A \textbf{78}, 042303 (2008).

\bibitem{Bell14b} B. A. Bell, D. A. Herrera-Mart\'i, M. S. Tame, D. Markham, W. J. Wadsworth, and J. G. Rarity, Experimental demonstration of a graph state quantum error-correction code, Nat. Commun. \textbf{5}, 3658 (2014).

\bibitem{Broadbent09} A. Broadbent, J. Fitzsimons, E. Kashefi, in \emph{Proceedings of the 50th Annual Symposium on Foundations of Computer Science} (IEEE Computer Society, Los Alamitos, CA, 2009), pp. 517--526.

\bibitem{Barz12} S. Barz, E. Kashefi, A. Broadbent, J. F. Fitzsimons, A. Zeilinger, and P. Walther, Demonstration of blind quantum computing, Science \textbf{20}, 303 (2012).

\bibitem{Giovannetti11} V. Giovannetti, S. Lloyd, and L. Maccone, Advances in quantum metrology, Nat. Photonics \textbf{5}, 222 (2011).

\bibitem{Komar14} P. K\'om\'ar E. M. Kessler, M. Bishof, L. Jiang, A. S. S\o rensen, J. Ye, and M. D. Lukin, A quantum network of clocks, Nat. Phys. \textbf{10}, 582--587 (2014).

\bibitem{Pirker17} A. Pirker, J. Walln\"ofer, and W. D\"ur, Modular architectures for quantum networks, New J. Phys. \textbf{20}, 053054 (2018).

\bibitem{Zhou03} D. L. Zhou, B. Zeng, Z. Xu, and C. P. Sun, Quantum computation based on d-level cluster state, Phys. Rev. A \textbf{68}, 062303 (2003).

\bibitem{He13} Q. Y. He and M. D. Reid, Genuine Multipartite Einstein-Podolsky-Rosen Steering, Phys. Rev. Lett. \textbf{111}, 250403 (2013).

\bibitem{Armstrong15} S. Armstrong, M. Wang, R. Y. Teh, Q. Gong, Q. He, J. Janousek, H.-A. Bachor, M. D. Reid, and P. K. Lam, Multipartite Einstein-Podolsky-Rosen steering and genuine tripartite entanglement with optical networks, Nat. Phys. \textbf{11}, 167 (2015).

\bibitem{Cavalcanti15} D. Cavalcanti, P. Skrzypczyk, G. H. Aguilar, R. V. Nery, P. H. Souto Ribeiro, and S. P. Walborn, Detection of entanglement in asymmetric quantum networks and multipartite quantum steering, Nat. Commun. \textbf{6}, 7941 (2015).

\bibitem{Deng17} X. Deng, Y. Xiang, C. Tian, G. Adesso, Q. He, Q. Gong, X. Su, C. Xie, and K. Peng, Demonstration of Monogamy Relations for Einstein-Podolsky-Rosen Steering in Gaussian Cluster States, Phys. Rev. Lett. \textbf{118}, 230501 (2017).

\bibitem{Li15} C.-M. Li, K. Chen, Y.-N. Chen, Q. Zhang, Y.-A. Chen, and J.-W. Pan, Genuine High-Order Einstein-Podolsky-Rosen Steering, Phys. Rev. Lett. \textbf{115}, 010402 (2015).

\bibitem{Xiang17} Y. Xiang, I. Kogias, G. Adesso, and Q. Y. He, Multipartite Gaussian steering: Monogamy constraints and quantum cryptography applications, Phys. Rev. A \textbf{95}, 010101(R) (2017).

\bibitem{Kogias17} I. Kogias, Y. Xiang, Q. Y. He, and G. Adesso, Unconditional security of entanglement-based continuous-variable quantum secret sharing, Phys. Rev. A \textbf{95}, 012315 (2017).

\bibitem{Monz11} T. Monz, P. Schindler, J. Barreiro, M. Chwalla, D. Nigg, W. Coish, M. Harlander, W. H\"ansel, M. Hennrich, and R. Blatt, 14-Qubit Entanglement: Creation and Coherence, Phys. Rev. Lett. \textbf{106}, 130506 (2011).

\bibitem{Barends14} R. Barends \textit{et al.}, Superconducting quantum circuits at the surface code threshold for fault tolerance, Nature (London) \textbf{508}, 500 (2014).

\bibitem{Kelly15} J. Kelly \textit{et al.}, State preservation by repetitive error detection in a superconducting quantum circuit, Nature (London) \textbf{519}, 66 (2015).

\bibitem{Wang16} X. L. Wang \textit{et al.}, Experimental Ten-Photon Entanglement, Phys. Rev. Lett. \textbf{117}, 210502 (2016).

\bibitem{Wang18} X. L. Wang \textit{et al.}, 18-qubit Entanglement with Six Photons's Three Degrees of Freedom, Phys. Rev. Lett. \textbf{120}, 260502 (2018).

\bibitem{Pappa12} A. Pappa, A. Chailloux, S. Wehner, E. Diamanti, and I. Kerenidis, Multipartite Entanglement Verification Resistant against Dishonest Parties, Phys. Rev. Lett. \textbf{108}, 260502 (2012). 

\bibitem{McCutcheon16} W. McCutcheon \textit{et al.}, Experimental verification of multipartite entanglement in quantum networks. Nat. Commun. \textbf{7}, 13251 (2016).

\bibitem{Li10} C.-M. Li, K. Chen, A. Reingruber, Y.-N. Chen, and J.-W. Pan, Verifying Genuine High-Order Entanglement, Phys. Rev. Lett. \textbf{105}, 210504 (2010).

\bibitem{Tanamoto06} T. Tanamoto, Y. X. Liu, S. Fujita, X. Hu, and F. Nori, Producing cluster states in charge qubits and flux qubits, Phys. Rev. Lett. \textbf{97}, 230501 (2006).

\bibitem{You07} J. Q. You, X. B. Wang, T. Tanamoto, and F. Nori, Efficient one-step generation of large cluster states with solid-state circuits, Phys. Rev. A \textbf{75}, 052319 (2007).

\bibitem{Tanamoto09} T. Tanamoto, Y. X. Liu, X. Hu, and F. Nori, Efficient Quantum Circuits for One-Way Quantum Computing, Phys. Rev. Lett. \textbf{102}, 100501 (2009).


\bibitem{Terhal2000} B. M. Terhal, and P. Horodecki, Schmidt number for density matrices, Phys. Rev. A \textbf{61}, 040301(R) (2000).

\bibitem{Tomamichel11} M. Tomamichel, and R. Renner, Uncertainty Relation for Smooth Entropies, Phys. Rev. Lett. \textbf{106}, 110506 (2011).

\bibitem{Coles17} P. J. Coles, M. Berta, M. Tomamichel, and S. Wehner, Entropic uncertainty relations and their applications, Rev. Mod. Phys. \textbf{89}, 015002 (2017).

\bibitem{Li15b} C.-M. Li, Y.-N. Chen, N. Lambert, C.-Y. Chiu, and F. Nori, Certifying single-system steering for quantum-information processing, Phys. Rev. A \textbf{92}, 062310 (2015).

\bibitem{Chiu16} C.-Y. Chiu, N. Lambert, T.-L. Liao, F. Nori, and C.-M. Li, No-cloning of quantum steering, npj Quantum Inf. \textbf{2}, 16020 (2016).


\bibitem{Buzek96} V. Bu\v{z}ek, and M. Hillery, Quantum copying: Beyond the no-cloning theorem, Phys. Rev. A \textbf{54}, 1844 (1996).

\bibitem{Scarani05} V. Scarani, S. Iblisdir, N. Gisin, and A. Ac\'{\i}n, Quantum cloning, Rev. Mod. Phys. \textbf{81}, 1225 (2005).

\bibitem{Sheridan10} L. Sheridan, and V. Scarani, Security proof for quantum key distribution using qudit systems, Phys. Rev. A \textbf{82}, 030301 (2010).

\bibitem{Zukowski98} M. \.{Z}ukowski, A. Zeilinger, M.A. Horne, and H. Weinfurter, Quest for GHZ states, Acta Phys. Pol. A \textbf{93}, 187 (1998).

\bibitem{DevetakWinter05} I. Devetak, and A. Winter, Distillation of secret key and entanglement from quantum states, Proceedings of the Royal Society of London A: Mathematical, Physical and Engineering Sciences \textbf{461}, 207 (2005).


\end{references}
\end{document}